\documentstyle[12pt,fleqn]{article}
\def\one{\mbox{{\sf 1}\kern-0.25em{\bf l}}}
\def\beq{\begin{equation}}
\def\eeq{\end{equation}}

\begin{document}
\newcommand{\av}[1]{\langle #1\rangle}

\vspace*{15mm}
\begin{center}
{\Large Bell Inequalities with Postselection$^*$} \vfill
Asher Peres\\[7mm]
{\sl Department of Physics, Technion -- Israel Institute of
Technology, 32 000 Haifa, Israel}\vfill

{\bf Abstract}\bigskip \end{center}

Experimental tests of Bell inequalities ought to take into account all
detection events. If the latter are postselected, and only some of
these events are included in the statistical analysis, a Bell
inequality may be violated, even by purely classical correlations. The
paradoxical properties of Werner states, recently pointed out by
Popescu, can be explained as the result of a postselection of the
detection events, or, equivalently, as due to the preparation of a new
state by means of a nonlocal process.

\vfill\noindent$^*$\,Dedicated to Professor Abner Shimony, on the
occasion of his 70th birthday.\vfill \newpage

Quantum mechanics is a statistical theory. It does not describe physical
processes that are happening in nature, but merely predicts {\it
probabilities of events\/}. Namely, if a physical system is prepared in
a definite way (mathematically represented by a Hermitian matrix
$\rho$), and that system is then subjected to a definite test
(represented by a projection operator $P$), the probability of passing
that test is equal to Tr$\,(\rho P)$. A natural question is whether
there is more to say than that. Can there be a more elaborate theory,
requiring a more detailed description of preparations and tests, such
that the outcomes of tests would be definite, rather than
probabilistic?

In 1964, John Bell proved that quantum mechanics is incompatible with
the existence of such a theory, if the latter has to obey the {\it
principle of local causes\/}. This principle (also called Einstein
locality, but conjectured well before Einstein) asserts that events
occuring in a given spacetime region are independent of external
parameters that may be controlled, at the same moment, by agents located
in distant spacetime regions. Bell's theorem [1] states that, as a
consequence of the principle of local causes, the outcomes of tests
performed on spatially distant systems cannot have arbitrarily large
correlations: the latter must satisfy a certain inequality. Bell
also showed that this inequality does not hold for spin correlations of
a pair of spin-$1\over2$ particles in a singlet state. That is, quantum
mechanics is incompatible with the principle of local causes.

At first, Bell's momentous discovery attracted only scant attention [2],
perhaps because in its original form, Bell's inequality had a restricted
domain of validity and could not be directly tested. However, in 1969,
Clauser, Horne, Shimony, and Holt [3] derived a more useful inequality,
valid under more general assumptions, and amenable to experimental
tests (it is called the CHSH inequality). Actual tests soon followed.
The most remarkable were those by Aspect and his collaborators [4, 5],
involving pairs of correlated photons originating from $SPS$ atomic
cascades.

Ironically, although the experiments fully agreed with the predictions
of quantum mechanics, doubts have been expressed whether a violation of
the CHSH inequality had actually been observed [6]. While no one denies
that the quantum mechanical formalism permits the existence of states
that violate the inequality, the interpretation of experimental results
is problematic: not all the particle detections are taken into account
in the statistical analysis. Some undesirable data are rejected. For
example, if only one of the two distant detectors is excited, the
unpaired detection event is ignored (this must often happen indeed,
since it is only occasionally that the two photons have nearly opposite
directions).

This selection of ``good'' data and rejection of ``bad'' ones is
suspicious. The use of biased statistical protocols is notorious for producing
outright fraud. In another quantum context, ``postselected'' measurement
data (namely, data sifted according to a procedure carried out {\it
after\/} completion of the measuring act) can sometimes yield average
values which are larger than the largest eigenvalue of an observable
[7]. To further illustrate this point, I shall now show how a simple
{\it classical\/} model can lead to a gross violation of the CHSH
inequality, if not all data are included in the statistical analysis.

Consider a massive classical object, initially at rest, which splits
into two parts carrying angular momenta {\bf J} and $-{\bf J}$. Let
${\bf n}={\bf J}/J$ be the unit vector in the direction of {\bf J}. The
direction of {\bf n} is random (it is isotropically distributed on the
unit sphere). Two distant observers, conventionally called Alice and
Bob, independently choose unit vectors {\bf a} and {\bf b},
respectively.  Alice measures ${\bf n\cdot a}$ and records a result,
$\alpha$, as follows:

\beq \begin{array}{lllll}
 \alpha=1 & & {\rm if} & & {\bf n\cdot a}>1/\sqrt{2},\\
 \alpha=-1 & & {\rm if} & & {\bf n\cdot a}<-1/\sqrt{2},\end{array} \eeq
and $\alpha=0$ in any other case. Likewise Bob measures $-{\bf n\cdot
b}$ and records $\beta=\pm1$ or 0, according to the same rule. (You can
easily visualize these rules by thinking of {\bf n} as the time axis in
a Minkowski spacetime. Then Alice records 1 or $-1$ when {\bf a} lies in
the future or past light cone, respectively, and Bob follows the same
rule for $-{\bf b}$.)

Obviously, Alice and Bob's results are correlated: they are controlled
by the common ``hidden'' variable {\bf n}. This is what Mermin [8] calls
a CLASS situation: {\bf C}orrelation {\bf L}ocally {\bf A}ttributable to
the {\bf S}ituation at the {\bf S}ource. (This term is meant to signify
a degree of virtue.)

It is easily shown that the correlation $\av{\alpha\beta}$ is a
continuous function of ${\bf a\cdot b}$, and takes values from
$\av{\alpha\beta}=1-2^{-1/2}\simeq0.3$, for ${\bf a\cdot b}=1$, to
$\av{\alpha\beta}\simeq-0.3$, for ${\bf a\cdot b}=-1$. If Alice and Bob
consider other possible testing directions, say ${\bf a}'$ and ${\bf
b}'$, and likewise define results $\alpha'$ and $\beta'$, the CHSH
inequality [3]

\beq  |\av{\alpha\beta}+\av{\alpha\beta'}
  +\av{\alpha'\beta}-\av{\alpha'\beta'}|\leq2, \label{CHSH} \eeq
is satisfied, as it should be for any CLASS model.

Suppose, however, that Alice and Bob consider a null result as a
failure, and retain in their statistics only those events where both
results differ from zero. It is easily seen that, in the events
postselected in that way, $\alpha\beta=1$ if the angle between
{\bf a} and {\bf b} is less than $90^\circ$ (this is a necessary
condition, not a sufficient one), and $\alpha\beta=-1$ if that
angle is more than $90^\circ$. Consider now four
directions, making angles of $45^\circ$, as shown in the figure.

\vfill
\hspace{10mm}\parbox{49mm}{The four directions used in
Eq.~(\ref{CHSH}) make angles of 45$^\circ$.}\vfill\newpage

We then have
\beq  \av{\alpha\beta}=\av{\alpha\beta'}=
  \av{\alpha'\beta}=-\av{\alpha'\beta'}=1, \eeq
and the left hand side of Eq.~(\ref{CHSH}) is equal to 4, so that the
CHSH inequality is grossly violated in this CLASS model. Even the
Cirel'son inequality [9, 10], which is respected by quantum mechanics,
is violated by the postselected results.

Why is this ridiculous example instructive? Some time ago, Werner [11]
constructed a density matrix $\rho_{\rm W}$ for a pair of spin-$j$
particles, with paradoxical properties. Werner's state $\rho_{\rm W}$
cannot be written as a sum of factorable density matrices, $\sum_j c_j\,
\rho_{{\rm A}j}\otimes\rho_{{\rm B}j}$, where $\rho_{{\rm A}j}$ and
$\rho_{{\rm B}j}$ belong to the two particles. Therefore, genuinely
quantal correlations are involved in $\rho_{\rm W}$.

For example, in the
simple case of a pair of spin-$1\over2$ particles, Werner's state is

\beq \rho_{\rm W}=\mbox{$1\over8$}\,\one+\mbox{$1\over2$}\,\rho_{\rm
  singlet}\,, \eeq
namely, an equal weight mixture of a totally uncorrelated random
state, and of a singlet state (the latter maximally violates the CHSH
inequality). A definitely nonclassical property of this $\rho_{\rm W}$
was discovered by Popescu [12], who showed that such a particle pair
could be used for teleportation of a quantum state [13], albeit with a
fidelity lesser than if a perfect singlet were employed for that purpose.

This nonclassical property is surprising, because, for any pair of
ideal local measurements performed on the two particles, the
correlations derived from $\rho_{\rm W}$ satisfy the CHSH inequality.
Moreover, as Werner showed, it is possible to introduce a
``hidden-variable'' model, which correctly produces all the observable
correlations for such ideal measurements. In this model, the hidden
variable is a unit vector {\bf r} in Hilbert space, and the quantum
probability rules are correctly reproduced if {\bf r} is isotropically
distributed.  Werner's prescription for the results of measurements of
projection operators is the following: if Alice considers a complete set
of orthonormal vectors ${\bf  v}_\mu$, and measures the corresponding
projection operators $P_\mu$, the result is $P_\mu=1$ for the ${\bf
v}_\mu$ having the smallest value of $|{\bf r\cdot v}_\mu|$ (that is,
the one most orthogonal to {\bf r}) and $P_\mu=0$ for all the other
${\bf v}_\mu$. For Bob, the rule is different and the results are only
probabilistic: the expectation value of $P_\mu$, for given {\bf r}, is
$\av{{\bf r},P_\mu{\bf r}}$.

Werner's algorithm for Alice's result becomes ambiguous for spin
$>\,{1\over2}$, and it must be supplemented by further rules, when we
consider projection operators of rank 2 or higher. For any projection
operator on a multi-dimensional subspace, Alice has to introduce, in an
arbitrary way, orthogonal frames which span that subspace and its
orthogonal complement. This defines a privileged complete orthogonal
basis, for which all the $P_\mu$ are defined as above. Then, the value
of a projection operator on any subspace is taken as equal to the sum of
the values, 0 or 1, of the projection operators on all the {\it
privileged\/} orthogonal vectors spanning that subspace. This rule is
unambiguous (once we have decided how to choose the privileged vectors),
but it has curious consequences.

Consider for example a 3-dimensional Hilbert space, with an orthogonal
basis \{{\bf x},{\bf y},{\bf z}\}. Let \{{\bf u},{\bf v},{\bf z}\}
be another orthogonal basis, so that \{{\bf x},{\bf y}\} and
\{{\bf u},{\bf v}\} span the same subspace, orthogonal to {\bf z}. Let
\{{\bf x},{\bf y}\} be our choice of privileged orthogonal basis for
defining the value of the projection operator, $P_{\bf xy}=P_{\bf uv}$,
on that subspace. It is then always possible to find ``hidden'' vectors
{\bf r} such that

\beq {\bf |r\cdot u|<|r\cdot z|<|r\cdot v|}, \eeq
and

\beq {\bf |r\cdot z|<|r\cdot x|<|r\cdot y|}. \eeq
In that case, Werner's rules imply that $P_{\bf u}=1$ if Alice
simultaneously measures $P_{\bf v}$ and $P_{\bf z}$, but, on the other
hand, the
value of $P_{\bf xy}=P_{\bf uv}$ is zero! This looks paradoxical, and
yet, {\it after averaging\/} over all {\bf r}, we still have

\beq \av{P_{\bf u}}+\av{P_{\bf v}}=\av{P_{\bf uv}}, \eeq
in agreement with quantum mechanics.

We thus see that the phrase ``a measurement of $P_{\bf u}$'' is
ambiguous.  We may have, for some values of the hidden variable {\bf r},
different outcomes depending on whether we measure $P_{\bf u}$ directly,
or we first perform a coarser measurement for $P_{\bf uv}$, which is
then refined for $P_{\bf u}$. This ambiguity was exploited by Popescu
[14], as follows. Instead of measuring complete sets of projection
operators of rank 1, Alice and Bob measure suitably chosen (and mutually
agreed) projection operators of rank 2, say $P_{\rm A}$ and $P_{\rm B}$.
If one of them gets a null result, the experiment is considered to have
failed, and they test another Werner pair. Only if both Alice and Bob
find the result 1, they proceed by independently choosing projection
operators of rank 1, on vectors that lie in the subspaces spanned by
$P_{\rm A}$ and $P_{\rm B}$, respectively. Popescu then shows that if
the initial Hilbert space (for each particle) has dimension 5 or higher,
the correlation of the final results violates the CHSH inequality. In
other words, Werner's hidden variable model, which worked for single
ideal measurements, is incapable of reproducing the results of several
{\it consecutive\/} measurements (and of course no other hidden variable
model would do it).

How can we understand this paradoxical result? We had what appeared to
be a CLASS model, similar to the classical model described at the
beginning of this essay. In the former case, the CHSH inequality was
violated as a result of faulty (postselected) statistics---all the
failures were discarded. The present case is subtler: Alice and Bob can,
if they wish, discard their failures {\it before\/} proceeding to the
final measurements. In other words, they can select a subensemble out of
the original ensemble, and it is this subensemble that violates the CHSH
inequality. The paradox is that the selection of this subensemble
apparently involves only {\it local\/} operations. How can it destroy
the CLASS property?

The point is that, in addition to the local measurements of $P_{\rm A}$
and $P_{\rm B}$, an exchange of classical information is needed for the
selection of the CHSH-violating subensemble. That classical information
is not just an abstract notion: it is conveyed by physical agents, such
as electro\-magnetic pulses. It is customary to consider information
carriers as \mbox{\it exophysical\/} systems [15], but this can only be
an approximation, which now raises suspicion. To further sharpen the
issue, let us promote the information carriers to \mbox{\it
endo\-physical\/} status, by attributing to them dynamical properties.
This leads to a new difficulty: there is no consistent hybrid dynamical
formalism for interacting classical and quantum systems. We must
therefore treat the information carriers as {\it quantum systems\/},
whose interaction with the Werner particles is generated by a
Hamiltonian, as usual. These additional quantum systems are manifestly
nonlocal, since their role is to propagate between Alice and Bob. It now
becomes obvious that the selection of the CHSH-violating subensemble
involves a {\it nonlocal\/} operation, and it is the latter that
violates the CLASS property of the original ensemble.

Popescu's construction [14] did not work for spaces with fewer than
$5^2$ dimensions, but similar protocols have been found [16--19] for
Werner pairs of spin-$1\over2$ particles. If each one of these pairs is
tested separately, the CHSH inequality is satisfied, as we know.  We
may, however, test several pairs together. For example, two pairs are
described by a $4^2$-dimensional space, in which there are nontrivial
rank-2 projection operators for each observer. Then, suitable
subensembles can be selected, that violate the CHSH inequality. It is
even possible to distill, from a large set of Werner pairs, a subset of
almost pure singlets [16--19]. Here again, no hidden variable model can
reproduce the results of {\it collective\/} measurements performed on
several Werner pairs.

In conclusion, we see that the notion of quantum nonlocality is subtler
than we may have thought. The conversion of a CLASS model into one that
violates the CHSH inequality can be explained in two equivalent ways: by
the use of biased statistics (postselected data), or by the introduction
of a nonlocal agent carrying information between the observers, before
completion of their measurements. Further investigations are needed, for
which the advice of Abner Shimony will be most precious.

\bigskip I am grateful to the authors of Refs.~[16--19] for advance
copies of their articles, and to N. D. Mermin and S. Popescu for an
illuminating exchange of correspondence. This work was supported by the
Gerard Swope Fund, and the Fund for Encouragement of Research.\clearpage

\frenchspacing
\begin{enumerate}
\item J. S. Bell, Physics 1 (1964) 195.
\item L. E. Ballentine, Am. J. Phys. 55 (1987) 785.
\item J. F. Clauser, M. A. Horne, A. Shimony, and R. A. Holt, Phys. Rev.
Lett. 23 (1969) 880.
\item A. Aspect, P. Grangier, and G. Roger, Phys. Rev. Lett. 47 (1981)
460; 49 (1982) 91.
\item A. Aspect, J. Dalibard, and G. Roger, Phys. Rev. Lett. 49 (1982)
1804.
\item E. Santos, Phys. Rev. Lett. 66 (1991) 1388.
\item Y. Aharonov, D. Z. Albert, and L. Vaidman, Phys. Rev. Lett. 60
(1988) 1351.
\item N. D. Mermin, unpublished Bielefeld lecture notes (1995).
\item B. S. Cirel'son, Lett. Math. Phys. 4 (1980) 93.
\item L. J. Landau, Phys. Lett. A120 (1987) 54.
\item R. F. Werner, Phys. Rev. A40 (1989) 4277.
\item S. Popescu, Phys. Rev. Lett. 72 (1994) 797.
\item C. H. Bennett, G. Brassard, C. Cr\'epeau, R.~Jozsa, A.~Peres, and
W.~K.~Wootters, Phys. Rev. Lett. 70 (1993) 1895.
\item S. Popescu, Phys. Rev. Lett. 74 (1995) 2619.
\item A. Peres, {\it Quantum Theory: Concepts and Methods\/} (Kluwer,
Dordrecht, 1993) p.~173.
\item C. H. Bennett, H. Bernstein, S. Popescu, and B. Schumacher, {\it
Concentrating Partial Entanglement by Local Operations\/} (submitted to
Phys. Rev. A, 1995).
\item C. H. Bennett, {\it Quantum and Clasical Information Transmission
Acts and Reducubilities\/} (to appear in proceedings of EPR60
conference, Haifa, 1995).
\item C. H. Bennett, G. Brassard, S. Popescu, B. Schumacher, J. Smolin,
and W. K. Wootters, {\it Purification of Noisy Entanglement, and
Faithful Teleportation via Noisy Channels\/} (submitted to Phys. Rev.
Lett., 1995).
\item D. Deutsch, A. K. Ekert, R. Jozsa, C. Macchiavello, S. Popescu,
and A. Sanpera, {\it Security of Quantum Cryptography over Noisy
Channels \/} (preprint, 1995).

\end{enumerate}

\end{document}